\documentclass[12pt,preprint]{aastex}

\newcommand{\sdssA}{SDSSp J$143952.58-003359.2$~}
\newcommand{\sdssB}{SDSSp J$143951.60-003429.2$~}
\newcommand{\lya}{Ly$\alpha$~}
\newcommand{\lyb}{Ly$\beta$~}

\newcommand{\vpar}{$v_{\|}$~}
\newcommand{\dvpar}{$\Delta v_{\|}$~}

\newcommand{\dtheta}{$\Delta\theta$~}
\newcommand{\kms}{km~s$^{-1}$}
\newcommand{\kmsMpc}{km~s$^{-1}$~Mpc$^{-1}$}
\newcommand{\hi}{H~{\sc i}~}
\newcommand{\civ}{C~{\sc iv}~}
\newcommand{\nv}{N~{\sc v}~}
\newcommand{\siii}{Si~{\sc ii}~}
\newcommand{\siiv}{Si~{\sc iv}~}
\newcommand{\mgii}{Mg~{\sc ii}~}
\newcommand{\feii}{Fe~{\sc ii}~}
\newcommand{\ovi}{O~{\sc vi}~}
\newcommand{\oi}{O~{\sc i}~}
\newcommand{\oif}{[O~{\sc i}]~}
\newcommand{\nai}{Na~{\sc i}~}
\newcommand{\Om}{$\Omega_m$}
\newcommand{\Ol}{$\Omega_{\Lambda}$}
\newcommand{\hs}{$h_{70}^{-1}$~}
\newcommand{\mF}{F}
\newcommand{\mdv}{\Delta v}
\newcommand{\mvpar}{v_{\|}}
\newcommand{\mdvpar}{\Delta v_{\|}}
\newcommand{\mdvper}{\Delta v_{\bot}}

\newcommand{\mdtheta}{\Delta\theta}

\newcommand{\mciv}{{\rm C \; \mbox{\tiny IV}}}

\newcommand{\mOm}{\Omega_m}
\newcommand{\mOl}{\Omega_{\Lambda}}

\newcommand{\AP}{Alcock-Paczy\'{n}ski}

\shorttitle{Correlations in the \lya forest}
\shortauthors{Becker, Sargent, & Rauch}

\begin{document}

\title{Large-Scale Correlations in the \lya Forest at $z=3-4$\altaffilmark{1}}

\author{George D. Becker\altaffilmark{2}, Wallace L. W. 
Sargent\altaffilmark{2}, Michael Rauch\altaffilmark{3}}

\altaffiltext{1}{The observations were made at the W.M. Keck Observatory
which is operated as a scientific partnership between the California
Institute of Technology and the University of California; it was made
possible by the generous support of the W.M. Keck Foundation.}
\altaffiltext{2}{Palomar Observatory, California Institute of Technology, 
Pasadena, CA 91125, USA; gdb@astro.caltech.edu, wws@astro.caltech.edu}
\altaffiltext{3}{Carnegie Observatories, 813 Santa Barbara Street, 
Pasadena, CA 91101, USA; mr@ociw.edu}

\begin{abstract}
We present a study of the spatial coherence of the intergalactic
medium toward two pairs of high-redshift quasars with moderate angular
separations observed with Keck/ESI, Q1422+2309A/Q1424+2255 ($z_{em}
\approx 3.63$, \dtheta = 39\arcsec) and Q1439-0034A/B ($z_{em} \approx
4.25$, \dtheta = 33\arcsec).  The crosscorrelation of transmitted flux
in the \lya forest shows a $5-7 \sigma$ peak at zero velocity lag for
both pairs.  This strongly suggests that at least some of the
absorbing structures span the $230-300$ \hs proper kpc transverse
separation between sightlines.  We also statistically examine the
similarity between paired spectra as a function of transmitted flux, a
measure which may be useful for comparison with numerical simulations.
In investigating the dependence of the correlation functions on
spectral characteristics, we find that photon noise has little impact
for S/N $\gtrsim 10$ per resolution element.  However, the agreement
between the autocorrelation along the line sight and the
crosscorrelation between sightlines, a potential test of cosmological
geometry, depends significantly on instrumental resolution.  Finally,
we present an inventory of metal lines.  These include a a pair of
strong \civ systems at $z \approx 3.4$ appearing only toward Q1439B,
and a \mgii + \feii system present toward Q1439 A and B at $z \approx
1.68$.
\end{abstract}


\section{Introduction}

Multiply-imaged lensed quasars and close quasar pairs provide valuable
probes of structure in the intergalactic medium.  By comparing the
absorption patterns in the spectra of adjacent quasar images one can
gauge the similarity in the underlying matter distributions along the
lines of sight and hence constrain the sizes of absorbing structures.
Previous studies at small separations ($\mdtheta \lesssim 10\arcsec$)
have typically provided lower limits to the scale of \lya absorbers
over a wide range in redshift, with weak upper constraints of $\sim
400$ comoving kpc derived by assuming spherical clouds
\citep{weymann83,foltz84,mcgill90,smette92,dinshaw94,bechtold94,
bechtold95,smette95,fang96}.  Observations of pairs at wider
separations ($\mdtheta \approx 0.5\arcmin-3\arcmin$) have shown
evidence that some \lya absorbers span $\gtrsim 1$ comoving Mpc
\citep{petitjean98,crotts98,dodorico98,dodorico02,young01,aracil02}
and possibly up to 30 comoving Mpc \citep{williger00}.

The majority of studies on lensed quasars and quasar pairs have relied
on matching individual absorption lines between spectra.  However,
high-resolution spectra (FWHM $\lesssim 20$ \kms) are required to
completely resolve these features.  In addition, the crowding of lines
in the high-redshift \lya forest often prevents the identification of
single absorbers.  An alternate approach reflecting the continuity of
the underlying density field is to compute the correlation of
transmitted flux along parallel lines of sight.  This robust statistic
is quickly computed and can be easily compared to numerical
simulations of large-scale structure.

As an extension to studying the matter distribution, correlations in
the \lya forest have been proposed as a tool for constraining the
cosmological constant through a variant of the \AP~test \citep{ap79}.
The \AP~test takes advantage of the fact that, for a homogeneous
sample of objects, the characteristic radial and transverse sizes
should be equal.  In the case of the \lya forest, one can compare the
correlation length of absorbing structures along the line of sight to
the correlation length in the transverse direction.  The velocity
separation, $\mdvpar$, between objects along the line of sight is
simply given by their redshifts,
\begin{equation}
\mdvpar = \frac{\Delta z}{1+z} c.  
\end{equation}
In contrast, the transverse velocity separation, $\mdvper$, between
objects at redshift $z$ with angular separation \dtheta depends on the
cosmological parameters implicit in the Hubble constant, $H(z)$, and
angular diameter distance, $D_A(z)$, as
\begin{equation}
\mdvper = H(z) \Delta l = H(z) D_A(z) \mdtheta,
\end{equation}
where $\Delta l$ is the proper linear separation.  One approach to
exploiting the difference between $\mdvpar$ and $\mdvper$ is to
directly compare the autocorrelation of transmitted flux along single
lines of sight to the crosscorrelation between spectra of sources at a
variety of angular separations \citep{hui99,lidz03}.  An alternate
method compares observed crosscorrelations to those determined from
linear theory \citep{mcdonald99} or from artificial spectra drawn from
numerical simulations \citep{lin03}.  Recently, \citet{rollinde03}
found agreement between the crosscorrelations and autocorrelations
among spectra of several $z \sim 2$ quasar pairs over a wide range in
separation (although see the discussion on spectral resolution below).
In the future, large surveys such as the Sloan Digital Sky Survey
\citep{sdssref} should greatly increase the number of quasar pairs
available for such studies.

We present results for two pairs of quasars at a novel combination of
high redshift and moderate separation.  The closely separated A and C
images ($\mdtheta = 1\farcs3$) of the bright $z=3.63$ lensed system
QSO 1422+2309 \citep{patnaik92} have been previously examined by
\citet{rauch99,rauch01} and \citet{rauch01b}.  \citet{adelberger03}
recently discovered an additional faint source, QSO 1424+2255
($z=3.62$) (therein referred to as Q1422b), at a separation $\mdtheta
= 38\farcs5$ from the lensed system.  In this study, we compare the
sightlines toward QSO 1422+2309A (herein referred to as Q1422) and
QSO 1424+2255 (herein referred to as Q1424 to avoid confusion with the
B image of Q1422).  We additionally investigate the $\mdtheta =
33\farcs4$ pair at $z=4.25$ discovered in the Sloan Digital Sky
Survey, \sdssA (herein Q1439A) and \sdssB (herein Q1439B)
\citep{schneider2000}.  For each pair, the similarity in quasar
redshifts allows us to study the transverse properties of the \lya
forest and intervening metal systems over a large pathlength.  Our
results should provide a valuable resource for comparison with
numerical simulations of large-scale structure.

The remainder of the paper is organized as follows: In \S 2 we present
our observations together with a general overview of the data.  We
compute the flux correlation functions in the \lya forest in \S 3 and
statistically examine the similarity between sightlines as a function
of flux.  In \S 4 we analyze the effects of photon noise and
instrumental resolution on the correlation functions.  Our results are
summarized in \S 5, with an inventory of unpublished metal absorption
systems presented in the appendix.

Throughout this paper we adopt $\mOm = 0.3$, $\mOl = 0.7$, and $H_0 =
70$ \kmsMpc.

\section{The Data}

Our observations are summarized in Table 1.  We observed all four
quasars under good to excellent seeing conditions over the period 2000
March to 2002 June using the Keck Echellette Spectrograph and Imager
(ESI) \citep{esiref} in echellette mode.  Additional observations of
Q1424 were provided by C. Steidel while additional observations of
Q1439A were provided by L. Hillenbrand.  All exposures were taken at
the parallactic angle except for one exposure of Q1439B at an airmass
near 1.0, where chromatic atmospheric dispersion is only a minor
concern.

The raw CCD frames for Q1422, Q1439A, and Q1439B were processed and
the 2-D echelle spectra extracted using the MAKEE software package.
Reduction of Q1424 data was performed using a suite of IRAF scripts,
as described in \citet{adelberger03}.  For each night we used the
extracted orders from at least one standard star (Feige 34, BD+284211,
and/or HZ 44) to determine an instrumental response function with
which to derive relative flux calibrations.  The calibrated orders of
all exposures from all nights were then converted to vacuum
heliocentric wavelengths and combined to produce a single continuous
spectrum per object (Figures 1 and 2).  We use a binned pixel size of
20 \kms.  Normalized spectra were produced by fitting continua to the
final reduced versions.

The majority of observations were made using a 0\farcs75 slit.
However, inspection of the spectra extracted from exposures taken with
different slits widths revealed very little difference in resolution,
likely due to favorable seeing.  For the combined spectra we adopt a
measured spectral resolution FWHM = 55 \kms.  The final ESI spectra
for Q1424, Q1439A, and Q1439B have typical S/N $\approx 13-30$ per
resolution element in the \lya forest, while Q1422 has S/N $\approx
150$.

We restrict our analysis of the \lya forest to the the wavelength
region between each quasar's \lya and \ovi emission lines.  To avoid a
proximity effect from the quasar, we include only those pixels at
least 10,000 \kms~blueward of the quasar's \lya emission.  We further
include only pixels at least 1,000 \kms~redward of \lyb and \ovi
emission to avoid any confusion with the \lyb forest or intrinsic \ovi
absorption.  For Q1422 and Q1424 this yields a redshift interval
$\Delta z = 0.50$ with mean redshift $\langle z \rangle = 3.22$.  For
Q1439 A and B, $\Delta z = 0.58$ and $\langle z \rangle = 3.79$.  Due
to the high redshift of Q1439 A and B, the \lya forest in the spectra
of these sources extends over the strong night sky lines \oif
$\lambda$5577 and \nai $\lambda$5890,5896.  In our analysis we exclude
the narrow ($2-3$ \AA) regions around these lines.

A visual comparison of the \lya forest in the paired spectra suggests
that the most striking similarity occurs among the strongest
absorption features (see Figures 3 and 4).  These regions often appear
to be similar along adjacent lines of sight, as do regions where the
absorption is nearly zero (possible ``voids'' relatively free of
absorbing material).  Matches among intermediate strength features are
less obvious, which suggests that the structures giving rise to those
lines may not be as coherent over the separation between sightlines.
However, many strong lines and regions of nearly zero absorption that
do not coincide between spectra can also be identified.  The alignment
of a subset of features may occur purely by chance.

Our spectral coverage also allows us to investigate the extent to
which metal systems span parallel lines of sight.  Two strong \civ
systems appear toward Q1439B at $z \approx 3.4$, separated by 1700
\kms.  Each system has a large rest equivalent width, $W_{\rm
rest}(1548) \sim 1.5$ \AA.  However, no \civ appears at this redshift
toward Q1439A.  In contrast, a low-ionization system containing \mgii
and \feii does appear along both sightlines at $z \approx 1.68$,
separated by only $\sim 400$ \kms.  A single, extended absorber
responsible for the low-ionization features would have a linear size
$\gtrsim 280$ \hs proper kpc.  These lines may alternatively arise
from the chance intersection of separate, clustered absorbers.  Weaker
\civ systems appear in both Q1422 and Q1424 separated by $\sim 440$
\kms~at $z \approx 3.08$, where the separation between sightlines is
290 \hs proper kpc.  We expand further on the properties of metal
systems in the appendix.

In order to assess the impact of spectral characteristics on our \lya
forest results we additionally employed a deep, high-resolution
spectrum of Q1422+2309A.  Observations were made with the Keck High
Resolution Echelle Spectrometer (HIRES) \citep{hiresref} using a
0\farcs574 slit.  This yields a spectral resolution FWHM = 4.4 \kms.
Reductions were performed as described \citet{rauch01}.  In addition
to the data from that work, which used the red cross-disperser only,
we include subsequent exposures taken with the UV-blazed
cross-disperser installed to provide additional coverage of the \lya
forest.  The final spectrum was binned to give a constant velocity
width for each pixel of 2.1 \kms, with a typical S/N per resolution
element of $50-90$.

\section{Comparison of Sightlines}

\subsection{Correlation Functions}

The correlation of transmitted flux in the spectra of closely
separated quasars provides a simple means of quantifying the degree of
similarity in the matter distribution along adjacent sightlines.  We
define the unnormalized correlation, $\xi$, of the spectra of two
sources separated on the sky by an angle $\mdtheta$ as
\begin{equation}
\xi(\mdtheta,\mdvpar) = \frac{1}{N} \sum
                  [\mF(\mdtheta,\mvpar + \mdvpar) - \bar{\mF}(\mdtheta)] 
		  [\mF(0,\mvpar) - \bar{\mF}(0)],
\end{equation}
where $\mF$ is the continuum-normalized flux, $\bar{\mF}(\mdtheta)$
and $\bar{\mF}(0)$ are the mean fluxes along the two sightlines, \vpar
is the line-of-sight velocity, \dvpar is the longitudinal velocity
lag, and $N$ is the total number of pixels in each spectrum within the
region of interest.  The sum is performed over all available pixels
at a given velocity lag, of which there will be
\begin{equation}
n_{\rm pix}(\mdvpar) = N - \mdvpar/\delta v_{\rm pix} 
\end{equation}
for a pixel size $\delta v_{\rm pix}$.  The normalized correlation
value can be computed by dividing Eq.~(3) by the standard deviation in
each input spectrum.

We have chosen a pixel size for the combined spectra to give roughly 3
pixels/resolution element.  However, we note that the $1/N$ factor in
Eq.~(3) implies that the value of the correlation will be insensitive
to pixel size so long as the spectrum is well sampled.  The increase
in the sum created by using a larger number of smaller pixels will be
offset by the increase in $N$ so long as there are $\gtrsim 2$
pixels/resolution element.  Using pixels larger than the spectral
resolution will introduce smoothing effects (see \S4).

In order to assess the coherence of absorbing structures across
adjacent sightlines, we compute the crosscorrelation of transmitted
flux in the \lya forest in the spectra of our quasar pairs.  As a
reference, we also determine the autocorrelation, $\xi(0,\mdvpar)$,
along lines of sight toward individual objects.  At these high
redshifts, \hi \lya absorption will strongly dominate over
contaminating absorption from lower-redshift metals such as \civ and
Mg~{\sc ii}.  The correlation functions should therefore accurately
reflect the distribution of neutral hydrogen to within the present
measurement errors.  All correlations are computed in single-pixel
steps, which are 20 \kms~for the ESI data.  Figure 5 displays the
autocorrelation function for each quasar along with the
crosscorrelation functions between adjacent sightlines.  Both in the
case of Q1422/Q1424 and Q1439A/B, a clear peak in the crosscorrelation
at zero lag indicates a genuine similarity between sightlines.

Undulations in the correlation functions arising from the chance
superposition of unrelated lines constitute the dominant source of
uncertainty in the peak values (see discussion on photon noise below).
Pixel-to-pixel variations in the correlations are themselves clearly
correlated.  However, we find that the overall distributions of values
away from the central peaks are very nearly Gaussian.  We therefore
take the standard deviation of pixels in the ``noise'' region, which
we define to be where 2000 \kms~$\leq \vert \mdvpar \vert \leq$ 18000
\kms, as the $1\sigma$ error in a correlation peak value.  In this
region we expect no underlying signal, however the correlation is
still computed from at least half of the available pixels.  Since the
sum in Eq.~(3) is computed over fewer pixels as the velocity lag
increases, yet the factor $1/N$ remains constant, the amplitude of the
noise features will tend to diminish as $\sqrt{n_{\rm pix}(\mdvpar)}$,
where $n_{\rm pix}(\mdvpar)$ is the number of pixels included in the
sum in Eq.~(3) at lag $\mdvpar$.  Therefore, in order to match the
amplitude of the noise at $\mdvpar=0$, we multiply the correlation at
each lag by a scale factor $s$ before computing the standard
deviation, where
\begin{equation}
s = \sqrt{\frac{n_{pix}(0)}{n_{\rm pix}(\mdvpar)}} 
 = \sqrt{\frac{N}{N-\mdvpar/\delta v_{\rm pix}}}.  
\end{equation}
For the range in velocity lag defined above this is a modest
correction.  Including only lags where at least half of the $\sim
1800$ \lya pixels in each spectrum overlap limits $s$ to at most
$\sqrt{2}$.  For the range in velocity lag shown in Fig. 5, $s < 1.2$.

Our results for the crosscorrelations are summarized in Table 2.
Using the above estimate for the error, the peak in the
crosscorrelation for Q1422/Q1424 (Q1439A/B) is significant at the
$7\sigma$ ($5\sigma$) level.  This strongly suggests coherence in the
absorbing structures on the scale of the $230-300$ \hs proper kpc
transverse separation between sightlines.  The marginal consistency of
the peak values with one another likely reflects the similarity in
sightline separation and redshift for the two quasar pairs.  The
zero-lag values of the autocorrelations, which give the variance in
the flux for these sections of the \lya forest, are 0.0950 for Q1422,
0.0796 for Q1424, 0.0862 for Q1439A, and 0.1000 for Q1439B.  Thus, the
normalized crosscorrelation peaks are $41.2 \pm 5.7\%$ for Q1422/Q1424
and $31.9 \pm 6.2\%$ for Q1439A/B.

\subsection{Flux Distributions}

The flux crosscorrelations demonstrate that at least some \lya
absorbers span the separation between our paired lines of sight.
However, it does not indicate whether the similarity between
sightlines depends on the strength of the absorber.  The crowded
nature of the \lya forest at $z \sim 4$, together with the present
spectral resolution, greatly inhibits a study of individual lines.
However, we are able to look at the agreement between absorption
features on a pixel-by-pixel basis.

Our goal is to determine whether the similarity in flux between paired
spectra depends on the amount of absorption for an individual pixel.
First we first select those \lya forest pixels in one spectrum (Q1422
or Q1439A) whose flux falls within a specified range.  We then
identify the pixels in the companion spectrum with matching
wavelengths and compute their flux distribution.  Comparing the
distribution in this subsample to that in all \lya forest pixels in
the companion spectrum allows us to evaluate whether there exists an
overabundance of pixels in the specified flux range relative to that
expected on random chance.

The results for Q1422/Q1424 and Q1439A/B are shown in Figures 6 and 7,
respectively.  Each panel shows the normalized distribution of Q1424
or Q1439B pixels in the indicated subsample along with the
distribution of all \lya forest pixels in that spectrum.  The range of
flux in Q1422 or Q1439A for each subsample is chosen to be
significantly larger than the typical flux uncertainty.  In each case,
we compute the two-sided Kolmogorov-Smirnov statistic, which is the
maximum difference between the cumulative fractions of pixels in the
subsample and of all pixels in the forest, along with the associated
likelihood of randomly obtaining a smaller statistic than the one
observed.  The number of pixels in each subsample is also shown.
 
The clearest results occur at extreme levels of absorption.  Pixels in
Q1422 (Q1439A) either near saturation, $F < 0.2$, or near the
continuum, $F > 0.8$, tend to strongly coincide with pixels of similar
flux in Q1424 (Q1439B).  Intermediate flux pixels appear to be less
strictly matched, with the exception of pixels in Q1422 with $0.4 < F
< 0.6$.  For certain subsamples, namely $0.6 < F < 0.8$ in Q1422 and
$0.4 < F < 0.6$ in Q1439A, the distribution of fluxes in the companion
spectrum is consistent with a random selection.  This seems to confirm
the visual appraisal that strong absorbers and regions relatively free
of absorbing material tend to span adjacent lines of sight more
readily than do absorbers of intermediate strength.

While each sample includes enough pixels to produce a statistically
significant result, a few caveats are worth considering.  Systematic
errors in continuum fitting might artificially create clusters of
pixels with low absorption near the same wavelength in pairs of
spectra.  Similarly, errors in sky-subtraction that are repeated
between spectra might create false coincidences of pixels near zero
flux.  Multiple independent continuum fits resulted in typical
differences of $\sim 10\%$ in flux on scales of $\sim 100$ \AA.  We
likewise find no evidence for large systematic errors in the sky
subtraction.  Given that the resulting uncertainties are significantly
smaller than the range in flux used to define a subsample in Figures 6
\& 7, these effects should only be a minor concern.

We stress that comparing transmitted fluxes does not strictly yield a
clear physical interpretation.  Pixels of intermediate flux commonly
occur along the wings of strong features.  They will therefore tend to
cluster less readily than pixels near the continuum or near
saturation, especially if the strong lines shift in velocity between
spectra.  Moreover, agreement in flux does not necessarily indicate
agreement in optical depth.  Since transmitted flux decreases
exponentially with optical depth, $\tau$, the difference in flux for a
given fractional change in $\tau$ will depend on the value of $\tau$
itself.  For a small characteristic change between sightlines, $\Delta
\tau/\tau \ll 1$, the greatest scatter in flux is expected for $\tau
\sim 1$, with the scatter decreasing as $\tau \rightarrow 0$ or as
$\tau \rightarrow \infty$.  Therefore, it is unclear whether the
enhanced agreement in flux among pixels with flux near the continuum
or near saturation indicates that the corresponding gas is more
homogeneous on these scales than the gas giving rise to intermediate
absorption features.  More specific insights may be drawn by comparing
our measured flux distributions to those derived from numerical
simulations.

\section{Effects of Photon Noise and Instrumental Resolution}

Comparing the longitudinal and transverse flux correlation functions
(or power spectra, equivalently) in the \lya forest has been been
explored by several authors as a means of measuring the cosmological
geometry via the \AP~test.  This application primarily requires a
sample of sightlines large enough to overcome cosmic variance.
However, some question remains regarding the dependence of the
correlation functions on data characteristics such as signal-to-noise
ratio and resolution.  It is particularly important to know to what
extent the autocorrelation measures the physical correlation length
along the line of sight rather than the instrumental resolution.  For
additional discussion on the effects of spectral characteristics see,
e.g., \citet{lin03} and \citet{croft98}.

In the preceding analysis we assumed that the chance alignment of
unrelated absorption features dominated over photon noise in producing
uncertainty in the flux correlations.  To justify this, we recomputed
the autocorrelation for the ESI spectrum of Q1422 after adding
increasing levels of Gaussian random noise.  The results appear nearly
identical for S/N $\gtrsim 10$ per resolution element (Figure 8).  A
spike at $\mdvpar = 0$ appears in the autocorrelation function as the
variance in the noise becomes comparable to the intrinsic variance in
the absorption features.  However, no such jump is expected to occur
in the crosscorrelation since the photon noise in the two spectra
should be uncorrelated.  Large sets of moderate S/N spectra may
therefore be more useful than smaller sets of high S/N for this type
of work.

To address the effects of spectral resolution on the autocorrelation
we have synthesized moderate- and low-resolution spectra from a
high-quality Keck HIRES spectrum of Q1422 (resolution FWHM = 4.4
\kms).  In each test case we first tune the spectral resolution by
convolving the HIRES data with a Gaussian kernel and then compute the
resulting autocorrelation function.  The results for spectra with
resolution FWHM = 4.4 (unsmoothed), 15, 55, and 200 \kms~are plotted
in Figure 9.  Very little difference exists between the
autocorrelation functions computed from the unsmoothed HIRES data and
from the data smoothed to FWHM = 15 \kms.  Thus, we may conclude that
the correlation in the unsmoothed HIRES data is an accurate measure of
the intrinsic correlation in the \lya forest (subject to cosmic
variance and redshift-space distortions).  The \lya lines are already
smoothed by their thermal width and easily resolved with HIRES.
However, when the spectrum is degraded to FWHM = 55 \kms, similar to
ESI data, the autocorrelation is clearly broadened and diminished in
amplitude.  (We note that the autocorrelation measured from this
``synthetic'' ESI spectrum is nearly identical to that computed from
the real ESI data.)  At even lower resolution, comparable to Keck/LRIS
or VLT/FORS2, the spectral resolution dominates over the intrinsic
correlation length.

As the spectral resolution decreases, the peak of the autocorrelation
also incorporates more of the outlying noise (random undulations in
the correlation function at $\mdv \gg 0$).  At large velocity lags
($\mdvpar \gg 200$ \kms), disagreement between the high-resolution and
low-resolution cases may be due in part to these noise features.
Better agreement may result when the correlation function is averaged
over many sightlines.

A more subtle issue is how spectral resolution affects the agreement
between autocorrelation and crosscorrelation functions.  To address
this, we degraded the resolution of our ESI data to mimic observations
using a lower resolution instrument (FWHM = 200 \kms) and then
recomputed the correlations.  Figure 10 contrasts the results from the
original spectra with those from the smoothed versions.  In each case
we plot the autocorrelation functions computed from single sightlines
and overplot the peak of the corresponding crosscorrelation at a
velocity lag equal to the transverse velocity separation between the
lines of sight ($\mdvper = 98$ \kms~for Q1422/Q1422, $\mdvper = 96$
\kms~for Q1439A/B for the case of \Om = 0.3, \Ol = 0.7, and $H_0 = 70$
\kmsMpc).

The differences shown in Figure 10 between the ESI and low-resolution
cases suggest a significant dependence on spectral resolution.  At ESI
resolution the concordance between auto- and crosscorrelations appears
to be very good for both pairs of sightlines.  However, at low
resolution the peaks of the crosscorrelation functions fall well below
the values of the autocorrelation functions at the same total velocity
separation.  While part of this effect may be due to noise in the
correlations, the general dependence on resolution can be understood
in regions where the intrinsic correlation function is non-linear on
scales smaller than the width of the smoothing kernel.  (This includes
the region around the peak at zero lag, since the correlation is
expected to be symmetric about $\mdvpar = 0$.)  Convolving the input
spectra with a smoothing kernel produces a correlation function that
has been convolved twice with the same kernel, once for each spectrum.
The resulting unnormalized autocorrelation function may therefore be
higher or lower at a particular velocity lag, depending on the shape
of the correlation function at that point.  However, since the peak of
the crosscorrelation function is already a maximum it can only
decrease as a result of smoothing (apart from the effects of noise).
Thus, if we adjust our cosmology such that, at high spectral
resolution, the composite crosscorrelation function built up from
pairs at many different separations agrees with the mean
autocorrelation function (ignoring redshift-space distortions), then
this agreement may not hold when using low-resolution data.  Spectral
resolution must therefore be considered carefully when performing this
type of comparison, for example by referring to artificial spectra
drawn from numerical simulations of large-scale structure.

\section{Conclusions}

We have analyzed the transverse properties of the \lya forest at $3.0
\lesssim z \lesssim 4.1$ on scales of $\sim 1$ \hs comoving Mpc
($230-300$ \hs proper kpc) toward the $z \approx 3.63$ quasar pair
Q1422+2309A/Q1424+2255 and the $z \approx 4.25$ pair Q1439-0034A/B.
Strong peaks at zero velocity lag in the flux crosscorrelations
between paired sightlines indicate coherence in the \hi absorbers on
these scales.  The crowded nature of the \lya forest at these
redshifts restricts a line-by-line comparison.  However, a statistical
approach suggests that the flux in paired spectra tends to be most
similar in regions of either very strong or very weak absorption.  The
similarities at flux levels near the continuum are consistent with the
large sizes of voids and low-density gaseous filaments and sheets
expected to comprise much of the intergalactic medium.  The more
surprising similarities at low flux levels may simply reflect the
limited dynamic range of transmitted flux in distinguishing among
regions of high optical depth ($\tau \gg 1$).  Larger differences in
flux found at intermediate flux levels may relate to the increased
scatter in flux naturally expected for optical depths near unity, and
also to the fact that flux differences along the wings of strong lines
will be sensitive to bulk velocity shifts between sightlines.  Despite
the difficulty in producing a clear physical interpretation from the
data alone, both the correlation and flux statistics should provide a
useful resource for comparison with numerical simulations.

In order to assess the dependence of the correlation functions on
photon noise and spectral resolution we have employed high-quality ESI
and HIRES spectra of Q1422.  Adding photon noise has very little
effect for S/N $\gtrsim 10$ per resolution element except to introduce
a spike in the autocorrelation at zero lag.  Increasing the resolution
FWHM increases the velocity width of the autocorrelation and decreases
its amplitude, as expected.  These effects may occur unevenly,
however, due to the incorporation of noise features as the peak
broadens.  We further tested the effects of resolution on the relative
amplitudes of the autocorrelation and crosscorrelation by degrading
the resolution of our ESI spectra to match that of a low-resolution
spectrograph such as Keck/LRIS or VLT/FORS2.  For both pairs, the
crosscorrelation and autocorrelations agree well for our adopted
cosmology when the correlations are computed from the original ESI
data.  However, smoothing the spectra substantially reduces the
crosscorrelation with respect to the autocorrelations.  This will
typically occur at any velocity lag where the intrinsic correlation
function is non-linear on scales smaller than the width of the
smoothing kernel.  Cosmological tests making use of the \lya forest
correlation length should take this effect into account.

Finally, we present an inventory of metal lines in Q1424, Q1439A, and
Q1439B in the appendix.  The most noteworthy features are a pair of
strong \civ complexes at $z \approx 3.4$ appearing only toward Q1439B.
Given their large equivalent widths, the relatively small separation
between these systems suggests that they are related, possibly arising
from separate galaxy halos or two sides of a galactic outflow.
Low-ionization systems containing \mgii and \feii appear at $z \approx
1.68$ in Q1439 A and B separated by 410 \kms.  These may either
represent a single, extended structure ($\Delta l \gtrsim 280$ \hs
proper kpc) or separate, clustered absorbers.  A \civ system appears
toward Q1424 at $z=3.084$, within 440 \kms~of a known \civ system
toward Q1422.  A single absorber responsible for both systems would
have a transverse size $\gtrsim 290$ \hs proper kpc.  However, the
number density of weak \civ systems at this redshift suggests that
these lines may instead result from chance superposition.

\acknowledgments

The authors would like to thank Chuck Steidel for bringing Q1424+2255
to our attention and for the additional data on this object, Kurt
Adelberger for his ESI reduction of Q1424, Lynne Hillenbrand for
kindly sharing her observations of Q1439A, and Rob Simcoe for his
HIRES reduction of Q1422A.  We would also like to thank Tom Barlow for
making public the MAKEE software package, Bob Carswell for VPFIT, and
the anonymous referee for helpful comments.  Finally, we are indebted
to the indigenous Hawaiian community for the opportunity to conduct
observations from their sacred mountain.  Without their hospitality
this work would not have been possible.  W.L.W.S gratefully
acknowledges support from the NSF through grants AST-9900733 and
AST-0206067.  M.R. has been supported by the NSF through grant
AST-0098492 and by NASA through grant AR-90213.01A.

\appendix

\section{Metal Systems}

Previous studies using quasar pairs and multiply-imaged lensed quasars
have demonstrated spatial coherence in metal systems over a variety of
length scales.  Differences between metal systems in
narrowly-separated lines of sight suggest that the absorbers giving
rise to individual lines seen in high-resolution spectra span at most
a few kiloparsecs, both for high-ionization systems seen in \civ and
for low-ionization systems seen in \mgii
\citep{lopez99,petitjean00,rauch01,rauch02,churchill03,tzanavaris03,ellison04}.
However, these absorbers may be part of larger structures extending
$\gtrsim 20$ \hs kpc \citep{smette95} and even $\gtrsim 100$ \hs kpc
for highly ionized material \citep{petitjean98,lopez99,lopez00}.

The present data afford us a unique opportunity to probe the coherence
of metal systems on scales of $\sim 300$ \hs proper kpc.
Unfortunately, modest signal-to-noise and significant contamination
from skylines in the red part of our spectra greatly limit our
sensitivity and hinder completeness estimates.  Our detections are
limited to relatively strong lines and lines that fall in regions of
unusually high S/N.

Line lists for Q1424, Q1439A, and Q1439B are presented in Tables 3, 4,
and 5, respectively.  In the following sections we briefly comment on
some of the more interesting systems.  Results for Q1422 are presented
in detail elsewhere \citep{rauch01,bechtold95,petry98}.  For each line
we measure the wavelength centroid and equivalent width.  Weighted
mean redshifts are given where multiple lines are measured for a
single ion.  In the case of blended lines we attempt to alleviate the
overlap by fitting Voigt profiles to any unblended transitions of the
same ions using VPFIT and then dividing by the inferred model
profiles.  This typically allows us to obtain values for at least the
strongest blended components.

\subsection{Q1424+2255}

Strong associated broad absorption lines (BALs) at $z = 3.62$ are seen
in \civ and \nv along with weaker associated absorption in \siiv
(although the \siiv appears affected by skyline contamination).
Isolated components of \civ and \nv appear up to 1000 \kms~blueward of
the BAL.  These suggest small pockets of outflowing material moving at
a higher velocity than the material responsible for the bulk of the
broad absorption.  \civ is also seen at $z = 3.62$ toward Q1422A.
\citep{rauch01}.  However, since the absorption is associated with the
quasar in both cases, these features are unlikely to be related.

An intervening \civ system appears at $z = 3.084$.  We additionally
identify the line at 5692.3 \AA~to be \siiv $\lambda$1394 at the same
redshift, where \siiv $\lambda$1403 is blended with \nv $\lambda$1239
at $z = 3.621$.  A \civ complex with possible \siiv appears at $z =
3.090$ toward both the A and C images of Q1422
\citep{rauch01,boksenberg03}, implying a transverse size $\gtrsim 290$
\hs proper kpc for a single structure spanning all three lines of
sight.  Figure 11 shows the \civ absorption at this redshift in Q1424
and Q1422.  The systems are narrowly separated along the line of
sight, with $\mdv \approx 440$ \kms.  However, \citet{boksenberg03}
find $\sim25$ \civ systems per unit redshift at $z = 3$ for systems
with total column density $N_{\rm syst}(\mciv) > 10^{12}$ cm$^{-2}$.
This translates to a typical spacing between systems of only $\sim
3000$ \kms.  It is therefore reasonable to suspect that this
superposition of relatively modest \civ absorbers toward Q1422 and
Q1424 occurs by chance.  Significant \hi absorption occurs along both
lines of sight in the velocity interval spanned by these systems,
though this may be due to the crowded state of the \lya forest at $z
\sim 3$.

\subsection{Q1439-0034A \& B}

Two broad \civ complexes at $z \approx 3.4$ form the most conspicuous
metal absorption features in the spectrum of Q1439B (Figure 12).  The
systems are separated by 1700 \kms, and each has a velocity width of
$\sim 400$ \kms.  Neither complex is observed toward Q1439A.  The
individual components are unlikely to be resolved with ESI.  However,
we can place a lower limit on the column density of these systems by
assuming that the weaker transition (\civ $\lambda$1551) falls on the
linear part of the curve-of-growth.  In that case, the column density,
$N_{\rm syst}$, is related to rest equivalent with, $W_{\rm rest}$, as
\begin{equation}
W_{\rm rest} = \frac{\pi e^2}{m_e c^2} f \lambda_0^2 N_{\rm syst},
\end{equation}
for a transition with rest wavelength $\lambda_0$ and oscillator
strength $f$.  For the lower-equivalent width system ($z_{\rm abs} =
3.402$, $W_{\rm rest}(1551) \approx 1.1$ \AA), this gives $N_{\rm
syst}(\mciv) \gtrsim 5 \times 10^{14}$ cm$^{-2}$.  If we adopt a
power-law column density distribution $d^2{\cal N}/dN_{\rm syst}dz
\propto N_{\rm syst}^{-\beta}$, where ${\cal N}$ is the number of
systems, then we would expect $\lesssim 0.6$ systems per unit redshift
in this column density range based on the above Boksenberg et al.\
(2003) results and adopting their value of $\beta = 1.6$.  The narrow
redshift spacing of these two systems ($\Delta z = 0.025$) therefore
strongly suggests that they are related.  One explanation is that the
absorption complexes arise from two sides of an intervening galactic
outflow.  Alternatively, the high levels of \hi absorption seen at
both \civ redshifts may indicate a pair of galaxy halos.

\mgii + \feii systems appear toward both Q1439 A and B at $z \approx
1.68$, separated between spectra by 410 \kms~along the line of sight
(Figure 13).  From \citet{churchill99}, the number of \mgii systems
per unit redshift with rest equivalent width $W_{\rm rest}(2796) >
0.02$ \AA~is expected to be $\approx 2.7$.  The probability of
randomly finding a system within this velocity separation is therefore
$\sim 0.02$, or a factor of three lower if we restrict ourselves to
strong ($W_{\rm rest}(2796) > 0.3$) systems \citep{steidel92}.  Thus,
there is a high likelihood that these systems are related.  Both \mgii
and \feii appear stronger toward B than toward A, with a factor of
seven difference between sightlines in the equivalent widths of \mgii
$\lambda$2804 and \feii $\lambda$2384.  These features may arise
either from a single absorber spanning the 280 \hs proper kpc between
sightlines or from a pair of clustered objects.  Two additional strong
\mgii systems appear along only a single sightline, one toward A at
$z=2.266$ and another toward B at $z=1.399$.

Q1439B is a BAL quasar showing high levels of associated absorption in
\civ and N~{\sc v}.  We identify the \oi absorption reported by
\citet{fukugita04} as the strong \civ systems at $z \approx 3.4$
mentioned above.  No strong associated absorption is seen for Q1439A.
As noted by \citet{fukugita04}, the difference in associated
absorption features supports the hypothesis that Q1439 A and B are a
true binary and not a lensed pair.

\clearpage

\begin{deluxetable}{llcc}
\tablecolumns{4}
\tabletypesize{\scriptsize}
\tablecaption{Summary of Observations}
\tablewidth{0pt}
\tablehead{
\colhead{Object} & \colhead{Date\tablenotemark{a}} & \colhead{Exp. time} & \colhead{FWHM\tablenotemark{b}} \\
 & & \colhead{(s)} & \colhead{(\kms)}  
} 
\startdata
\cutinhead{Keck/ESI}
Q1422+2309A & 2000 Mar 03-04             & 1200\tablenotemark{c}  & $\sim 55$ \\
Q1424+2255  & 2000 Mar 03 -- 2001 Apr 21 & 44900\tablenotemark{d} & $\sim 55$ \\ 
Q1439-0034A & 2001 Jan 26 -- 2002 Jun 10 & 8000\tablenotemark{e}  & $\sim 55$ \\ 
Q1439-0034B & 2001 Apr 19 -- 2002 Jun 10 & 42400\tablenotemark{f} & $\sim 55$ \\ 
\cutinhead{Keck/HIRES}
Q1422+2309A & 1998 Jan 30 -- 1998 Apr 15 & 31600\tablenotemark{g} & 4.4 \\
\enddata
\tablenotetext{a}{Interval over which observations were made.}
\tablenotetext{b}{Spectral resolution.}
\tablenotetext{c}{600 s with 0\farcs75 slit, 600 s with 1\farcs0 slit.}
\tablenotetext{d}{43100 s with 0\farcs75 slit (including 20900 s by C. Steidel), 1800 s with 1\farcs0 slit.}
\tablenotetext{e}{5000 s with 0\farcs75 slit, 3000 s with 0\farcs5 slit (L. Hillenbrand).}
\tablenotetext{f}{0\farcs75 slit.}
\tablenotetext{g}{9000 s with UV cross-disperser, 22600 s with red cross-disperser}  
\end{deluxetable}

\begin{deluxetable}{lcccccc}
\tabletypesize{\scriptsize}
\tablecaption{\lya Forest Crosscorrelation Values}
\tablewidth{0pt}
\tablehead{
\colhead{QSO Pair} & \colhead{$\mdtheta$\tablenotemark{a}} & \colhead{$z_{{\rm Ly}\alpha}$\tablenotemark{b}} 
& \colhead{$\Delta l$\tablenotemark{c}} & \colhead{$\mdvper$\tablenotemark{d}} & \multicolumn{2}{c}{$\xi(\mdtheta,0)$} \\
\cline{6-7}
 & & & \colhead{\hs proper kpc} & \colhead{\kms} & \colhead{Unnormalized\tablenotemark{e}} & \colhead{Normalized\tablenotemark{f}} 
}
\startdata
Q1422/Q1424 & $38\farcs5$ & $2.965-3.466$ & $283-298$ & 97.8 & $0.0358 \pm 0.0049$ & $0.412 \pm 0.057$ \\
Q1439A/B    & $33\farcs4$ & $3.496-4.075$ & $230-245$ & 96.3 & $0.0296 \pm 0.0057$ & $0.319 \pm 0.062$ \\
\enddata
\tablenotetext{a}{Angular separation of the QSO pair.}
\tablenotetext{b}{Redshift interval used in computing the correlation values for the \lya forest.}
\tablenotetext{c}{Linear separation between the lines of sight for the given redshift interval, computed for 
$\mOm = 0.3$, $\mOl = 0.7$, and $H_0 = 70$ \kmsMpc.}
\tablenotetext{d}{Transverse velocity separation between the lines of sight for our adopted cosmology 
at the mean redshift in the given interval.}
\tablenotetext{e}{Crosscorrelation at zero lag computed from Eq.~(3).}
\tablenotetext{f}{Crosscorrelation at zero lag computed from Eq.~(3) divided by the standard deviation 
of the \lya forest flux in each input spectrum.}
\end{deluxetable}

\begin{deluxetable}{lccc}
\tablecolumns{4}
\tabletypesize{\scriptsize}
\tablecaption{Q1424+2255 Metal Systems \label{q1424metals}}
\tablewidth{0pt}
\tablehead{
\colhead{$\lambda_{\rm obs}$} & \colhead{Line ID} & \colhead{$z_{\rm obs}$\tablenotemark{a}} & \colhead{$W_{\rm rest}$} \\
\colhead{(\AA)} & & & \colhead{(\AA)}
}
\startdata
\cutinhead{Intervening}
$5692.34 \pm 0.12$ & \siiv $\lambda$1394 & $3.08417 \pm 0.00009$ & $0.29 \pm 0.02$ \\
$6323.34 \pm 0.14$ & \civ $\lambda$1548 & $3.08433 \pm 0.00007$ & $0.50 \pm 0.03$ \\
$6333.93 \pm 0.17$ & \civ $\lambda$1550 & $3.08433 \pm 0.00007$ & $0.37 \pm 0.03$ \\
\cutinhead{Associated}
$5704.76 \pm 0.08$ & \nv $\lambda$1239 & $3.60499 \pm 0.00007$ & $0.34 \pm 0.01$ \\
$5709.70 \pm 0.11$ & \nv $\lambda$1239 & $3.60898 \pm 0.00009$ & $0.11 \pm 0.01$ \\
$5724.81 \pm 0.08$ & \nv $\lambda$1239 & $3.62118 \pm 0.00005$ & $0.96 \pm 0.02$ \\
$5743.22 \pm 0.09$ & \nv $\lambda$1243 & $3.62118 \pm 0.00005$ & $0.75 \pm 0.02$ \\
$6437.34 \pm 0.17$ & \siiv $\lambda$1394 & $3.61865 \pm 0.00009$ & $0.24 \pm 0.03$ \\
$6478.79 \pm 0.21$ & \siiv $\lambda$1403 & $3.61865 \pm 0.00009$ & $0.16 \pm 0.02$ \\
$7129.64 \pm 0.08$ & \civ $\lambda$1548 & $3.60511 \pm 0.00005$\tablenotemark{b} & $0.36 \pm 0.01$ \\
$7135.69 \pm 0.16$ & \civ $\lambda$1548 & $3.60902 \pm 0.00010$ & $0.10 \pm 0.01$ \\
$7141.28 \pm 0.08$ & \civ $\lambda$1550 & $3.60511 \pm 0.00005$\tablenotemark{b} & $\lesssim 0.37$\tablenotemark{c} \\
$7144.70 \pm 0.13$ & \civ $\lambda$1548 & $3.61484 \pm 0.00008$ & $0.08 \pm 0.01$ \\
$7153.99 \pm 0.07$ & \civ $\lambda$1548 & $3.62085 \pm 0.00004$ & $1.82 \pm 0.03$ \\
$7165.96 \pm 0.11$ & \civ $\lambda$1550 & $3.62085 \pm 0.00004$ & $1.44 \pm 0.06$ \\
\enddata
\tablenotetext{a}{Weighted mean redshift from all available transitions.}
\tablenotetext{b}{\civ doublet redshift computed from the $\lambda$1548 line only.}
\tablenotetext{c}{3$\sigma$ upper limit (possible blend).}
\end{deluxetable}

\begin{deluxetable}{lccc}
\tablecolumns{4}
\tabletypesize{\scriptsize}
\tablecaption{Q1439-0034A Metal Systems \label{q1439ametals}}
\tablewidth{0pt}
\tablehead{
\colhead{$\lambda_{\rm obs}$} & \colhead{Line ID} & \colhead{$z_{\rm obs}$\tablenotemark{a}} & \colhead{$W_{\rm rest}$} \\
\colhead{(\AA)} & & & \colhead{(\AA)}
}
\startdata
\cutinhead{Intervening}
$6393.06 \pm 0.10$ & \feii $\lambda$2383 & $1.68304 \pm 0.00004$ & $0.07 \pm 0.01$ \\
$7502.82 \pm 0.12$ & \mgii $\lambda$2796 & $1.68305 \pm 0.00004$ & $0.18 \pm 0.02$ \\
$7521.87 \pm 0.19$ & \mgii $\lambda$2804 & $1.68305 \pm 0.00004$ & $0.10 \pm 0.02$ \\
$9134.29 \pm 0.14$ & \mgii $\lambda$2796 & $2.26645 \pm 0.00004$ & $0.30 \pm 0.03$ \\
$9157.37 \pm 0.18$ & \mgii $\lambda$2804 & $2.26645 \pm 0.00004$ & $0.32 \pm 0.04$ \\
\enddata
\tablenotetext{a}{Weighted mean redshift from all available transitions.}
\end{deluxetable}

\begin{deluxetable}{lccc}
\tablecolumns{4}
\tabletypesize{\scriptsize}
\tablecaption{Q1439-0034B Metal Systems \label{q1439bmetals}}
\tablewidth{0pt}
\tablehead{
\colhead{$\lambda_{\rm obs}$} & \colhead{Line ID} & \colhead{$z_{\rm obs}$\tablenotemark{a}} & \colhead{$W_{\rm rest}$} \\
\colhead{(\AA)} & & & \colhead{(\AA)}
}
\startdata
\cutinhead{Intervening}
$6402.03 \pm 0.06$ & \feii $\lambda$2382 & $1.68680 \pm 0.00002$ & $0.51 \pm 0.01$ \\
$6602.02 \pm 0.13$ & \civ $\lambda$1548 & $3.26419 \pm 0.00006$ & $0.15 \pm 0.02$ \\
$6612.57 \pm 0.14$ & \civ $\lambda$1551 & $3.26419 \pm 0.00006$ & $0.13 \pm 0.02$ \\
$6704.31 \pm 0.11$ & \civ $\lambda$1548 & $3.33041 \pm 0.00006$ & $0.14 \pm 0.02$ \\
$6707.40 \pm 0.06$ & \mgii $\lambda$2796 & $1.39863 \pm 0.00002$ & $1.18 \pm 0.03$ \\
$6709.91 \pm 0.11$ & \siii $\lambda$1527 & $3.39502 \pm 0.00007$ & $0.17 \pm 0.02$ \\
$6715.59 \pm 0.19$ & \civ $\lambda$1551 & $3.33041 \pm 0.00006$ & $0.09 \pm 0.02$ \\
$6724.68 \pm 0.07$ & \mgii $\lambda$2804 & $1.39863 \pm 0.00002$ & $0.91 \pm 0.03$ \\
$6804.02 \pm 0.07$ & \civ $\lambda$1548 & $3.39476 \pm 0.00003$ & $1.48 \pm 0.02$ \\
$6815.18 \pm 0.08$ & \civ $\lambda$1551 & $3.39476 \pm 0.00003$ & $1.09 \pm 0.02$ \\
$6843.29 \pm 0.06$ & \civ $\lambda$1548 & $3.42022 \pm 0.00003$ & $1.58 \pm 0.04$ \\
$6854.89 \pm 0.06$ & \civ $\lambda$1551 & $3.42022 \pm 0.00003$ & $1.25 \pm 0.02$ \\
$6949.75 \pm 0.20$ & \feii $\lambda$2587 & $1.68680 \pm 0.00002$ & $0.26 \pm 0.05$ \\
$6986.15 \pm 0.08$ & \feii $\lambda$2600 & $1.68680 \pm 0.00002$ & $0.50 \pm 0.02$ \\
$7387.85 \pm 0.14$ & \civ $\lambda$1548 & $3.77191 \pm 0.00007$ & $0.18 \pm 0.02$ \\
$7400.22 \pm 0.19$ & \civ $\lambda$1551 & $3.77191 \pm 0.00007$ & $0.09 \pm 0.02$ \\
$7513.04 \pm 0.08$ & \mgii $\lambda$2796 & $1.68673 \pm 0.00003$\tablenotemark{b} & $0.70 \pm 0.03$ \\
$7532.51 \pm 0.15$ & \mgii $\lambda$2804 & $1.68673 \pm 0.00003$\tablenotemark{b} & $0.68 \pm 0.13$ \\
\cutinhead{Associated}
$6502.00 \pm 0.09$ & \nv $\lambda$1239 & $4.24877 \pm 0.00005$ & $1.62 \pm 0.05$ \\
$6523.45 \pm 0.08$ & \nv $\lambda$1243 & $4.24877 \pm 0.00005$ & $1.31 \pm 0.02$ \\
$8125.49 \pm 0.04$ & \civ $\lambda$1548 & $4.24847 \pm 0.00002$ & $1.94 \pm 0.01$ \\
$8139.44 \pm 0.04$ & \civ $\lambda$1551 & $4.24847 \pm 0.00002$ & $1.68 \pm 0.01$ \\
\enddata
\tablenotetext{a}{Weighted mean redshift from all available transitions.}
\tablenotetext{b}{\mgii doublet redshift computed from the $\lambda$2796 line only.}
\end{deluxetable}

\clearpage

\begin{figure} 
\epsscale{1} 
\plotone{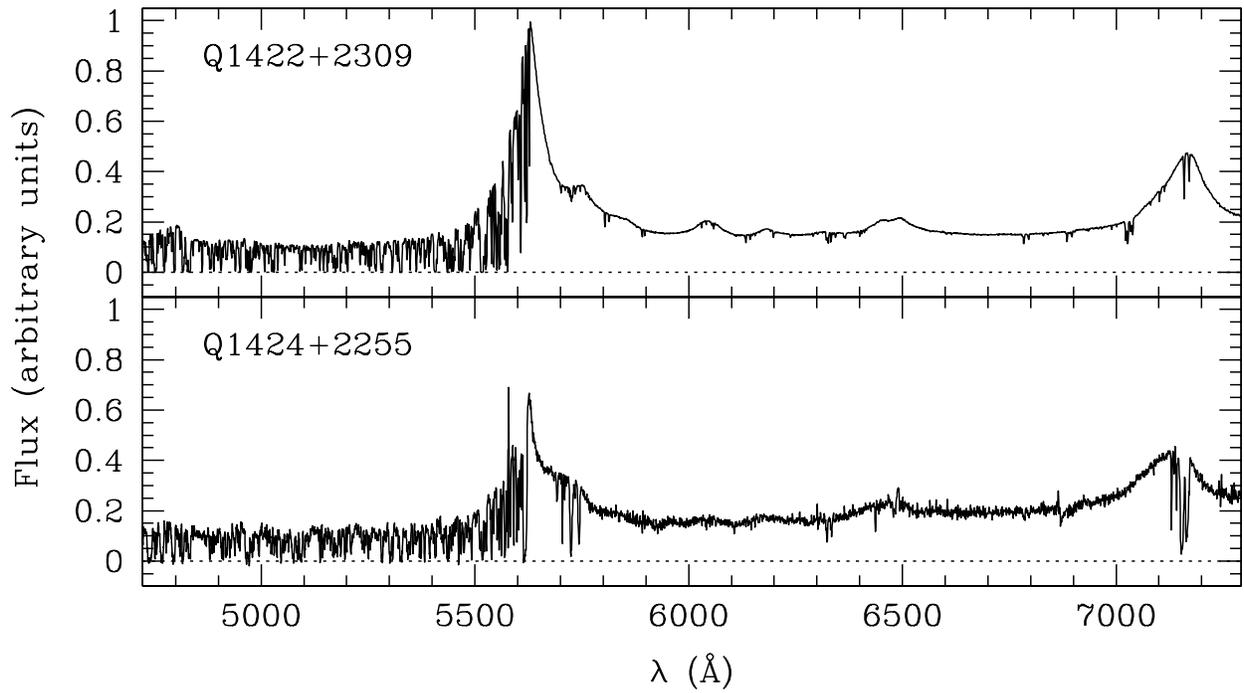} 
\figcaption{Keck/ESI spectra covering the rest wavelength region from 
\lyb to \civ in Q1422 ({\it top}) and Q1424 ({\it bottom}).  Spectra have 
been binned in wavelength using 40 \kms~pixels for display.  Contamination 
from the \oif $\lambda$5577 skyline can be seen in the spectrum of Q1424.}
\label{q1422_full}
\end{figure}

\clearpage

\begin{figure} 
\epsscale{1} 
\plotone{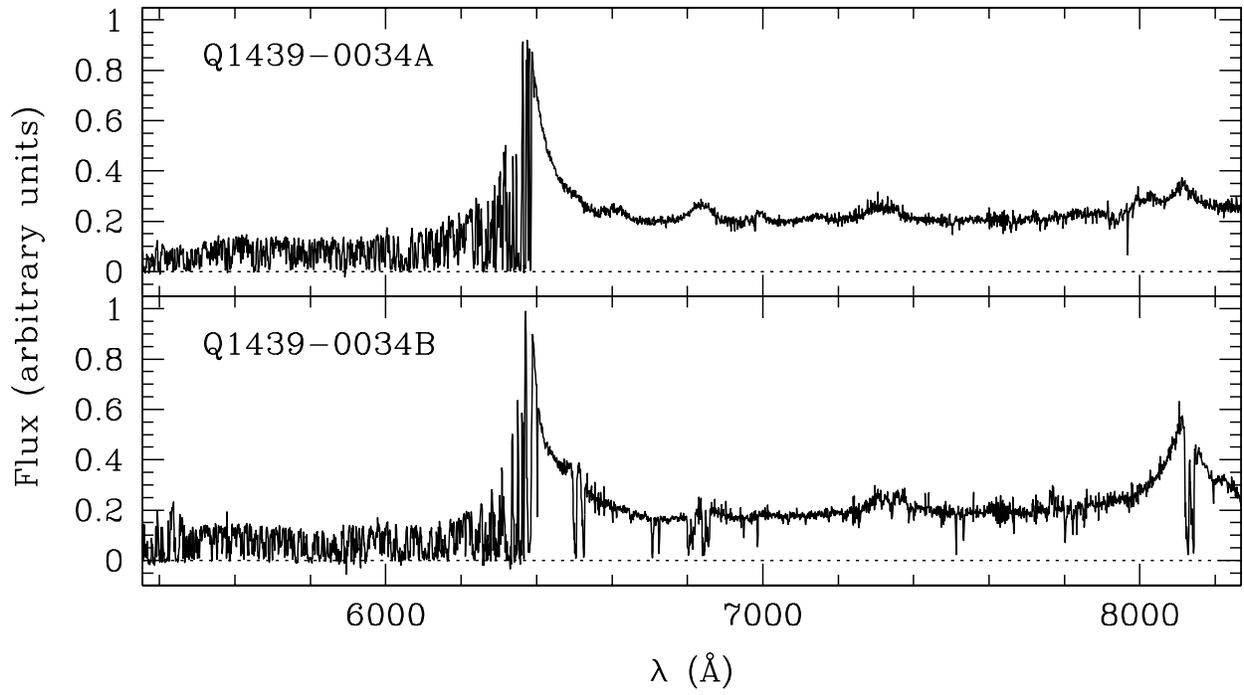} 
\figcaption{Same as Figure 2, here for Q1439A ({\it top}) and Q1439B 
({\it bottom}).}
\label{q1439_full}
\end{figure}

\clearpage

\begin{figure}
\epsscale{1} 
\plotone{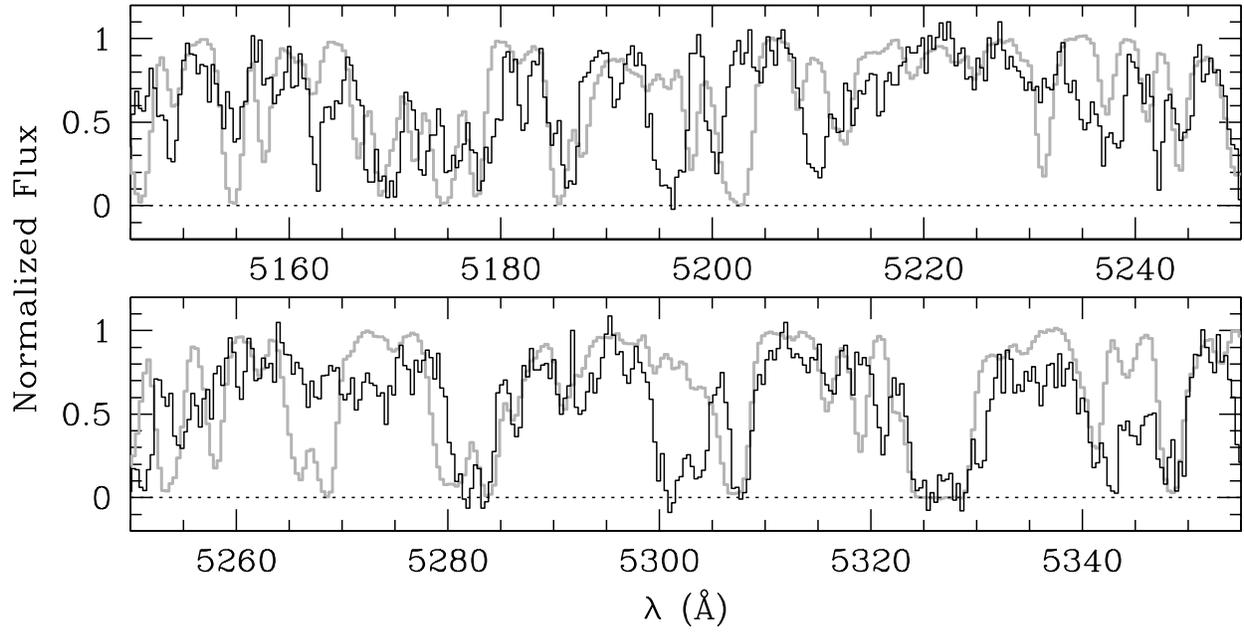} 
\figcaption{A representative section of the \lya forest in the normalized 
spectra of Q1422 (grey line) and Q1424 (black line).}
\label{q1422_lya}
\end{figure}

\clearpage

\begin{figure}
\epsscale{1} 
\plotone{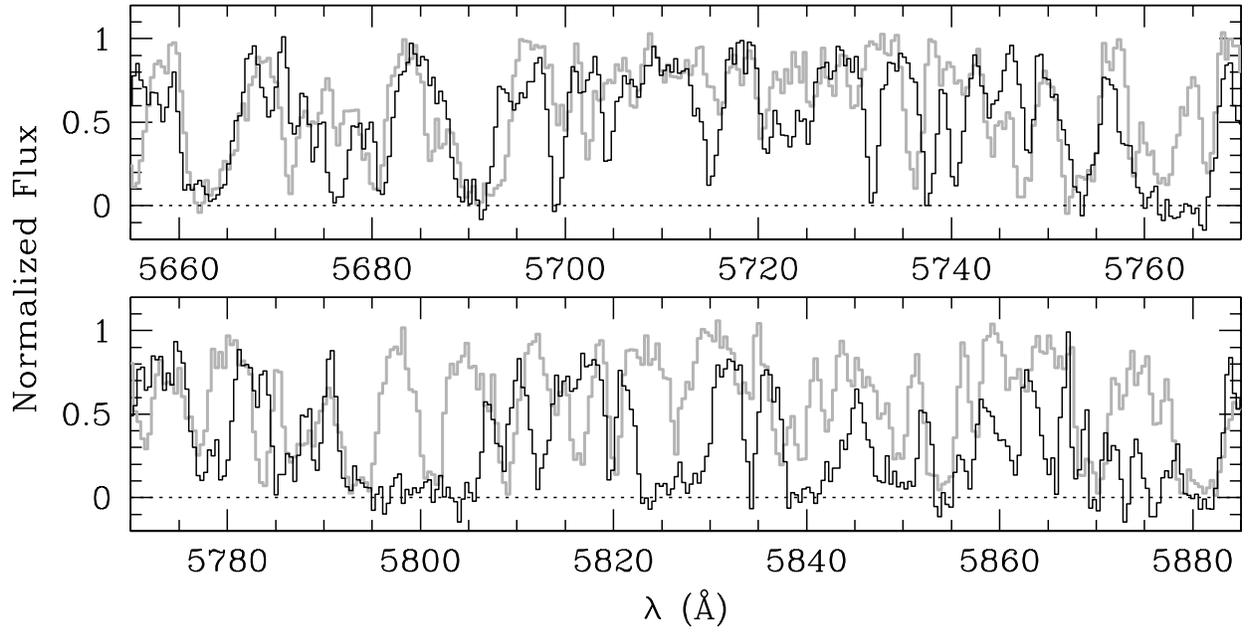} 
\figcaption{Same as Figure 3, here for Q1439A (grey line) and Q1439B (black
line).}
\label{q1439_lya}
\end{figure}

\clearpage

\begin{figure}
\epsscale{1} 
\plotone{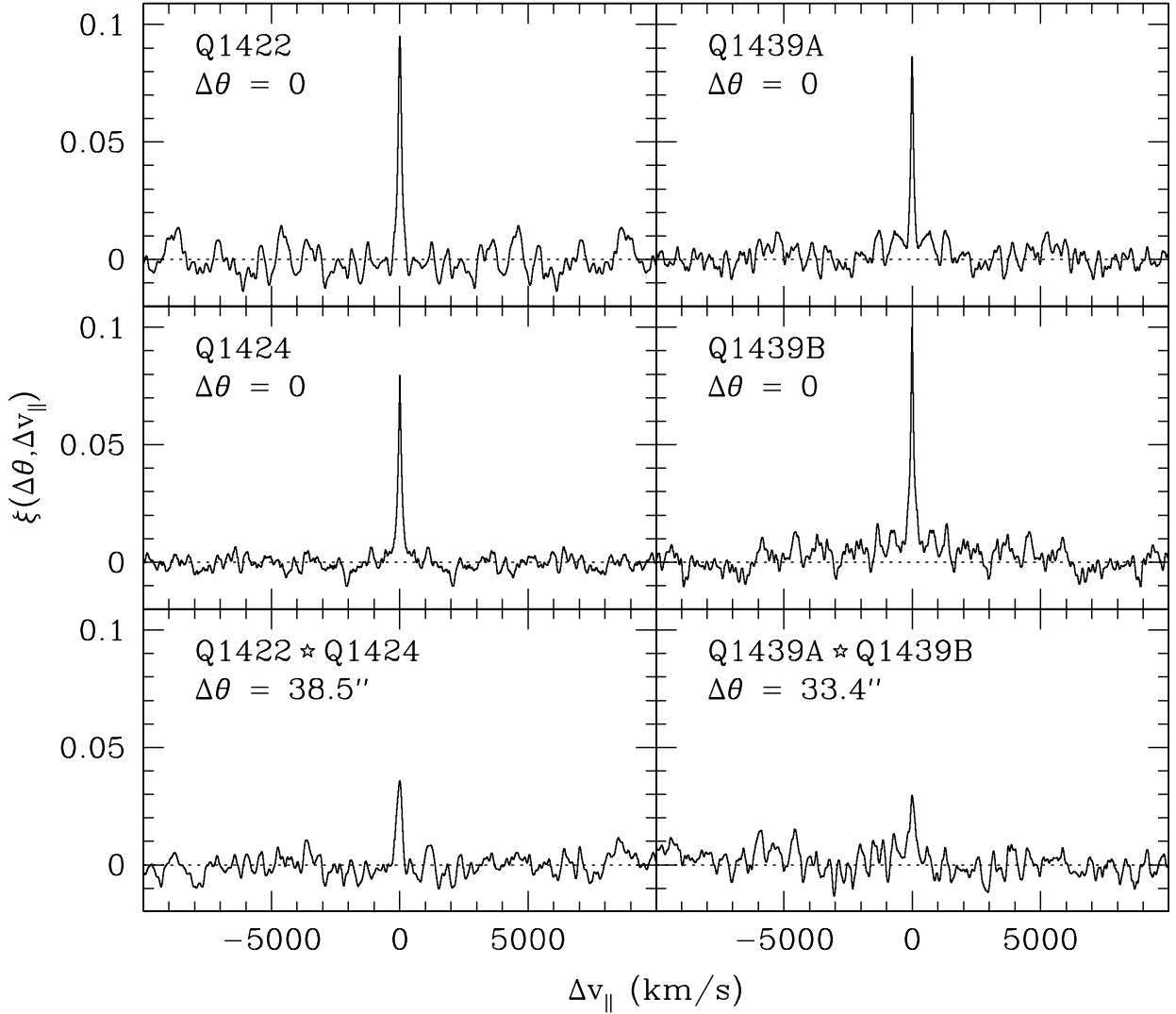} 

\figcaption{Flux correlation functions computed from the \lya forest
plotted vs. longitudinal velocity lag.  {\it Top four panels:}
Autocorrelation functions for Q1422 ({\it top left}), Q1424 ({\it
middle left}), Q1439A ({\it top right}), and Q1439B ({\it middle
right}).  Negative velocity lags have been included for consistency.
{\it Bottom panels:} Crosscorrelation functions for Q1422/Q1422 ({\it
bottom left}) and Q1439A/B ({\it bottom right}).}

\label{cross_cov_comb}
\end{figure}

\clearpage

\begin{figure}
\epsscale{1} 
\plotone{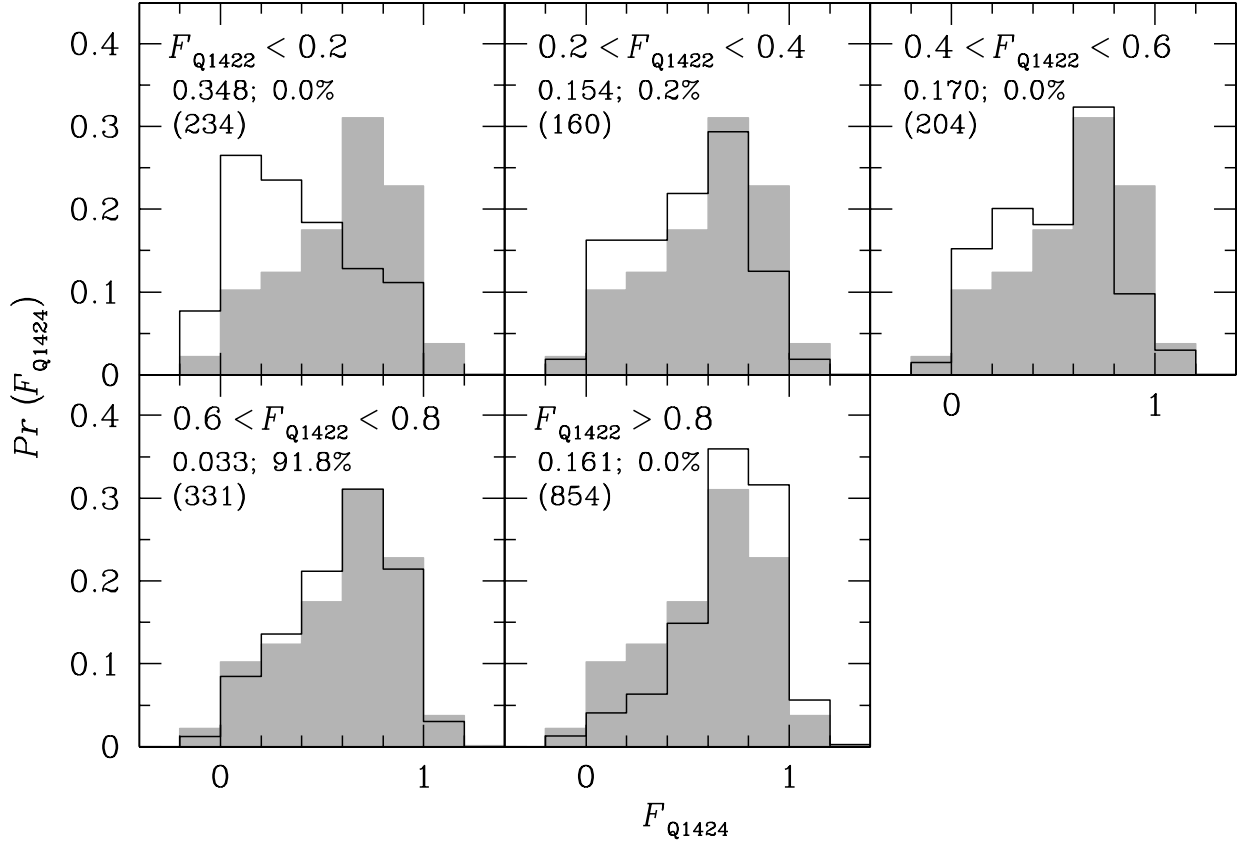} 
\figcaption{Distributions of continuum-normalized pixel fluxes in the
\lya region of Q1424 as a function of the flux in Q1422.  Unshaded
histograms show the flux probability distribution for the subsample of
pixels in Q1424 corresponding in wavelength to those pixels in Q1422
that have fluxes in the indicated range.  Shaded histograms show the
probability distribution for all Q1424 pixels in the \lya region.  For
each range of flux in Q1422, the two-sided Kolmogorov-Smirnov
statistic for the subsample and full sample of Q1424 flux values is
presented along with the associated likelihood, expressed as a
percent, of obtaining a higher value if the the two samples were drawn
from the same distribution.  The number of pixels in each subsample is
given in parentheses.  The full sample contains 1783 pixels.}
\label{fluxpdf_1422}
\end{figure}

\clearpage

\begin{figure}
\epsscale{1} 
\plotone{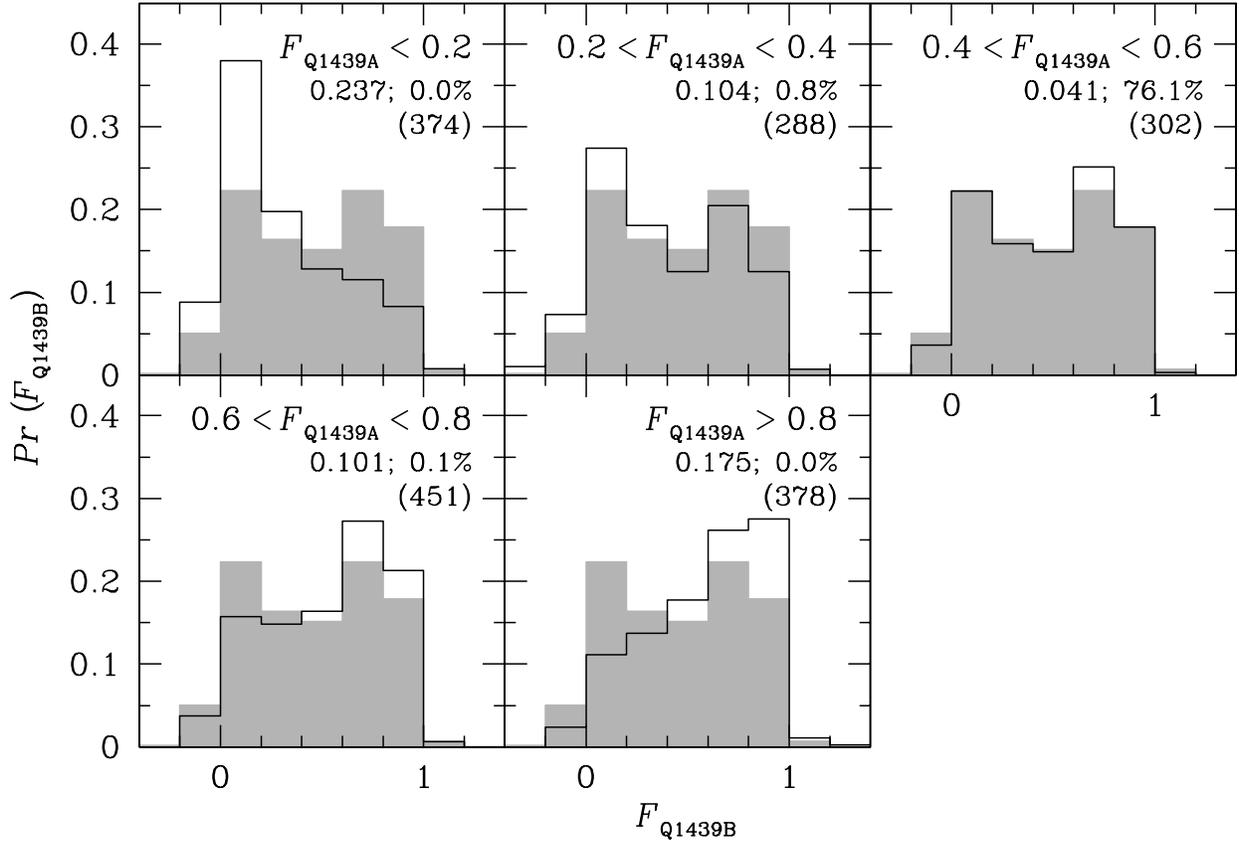} 
\figcaption{Same as Figure 6, here for pixel fluxes in the \lya region 
of Q1439B as a function of the flux in Q1439A.  The full sample contains
1793 pixels.}
\label{fluxpdf_1439}
\end{figure}

\clearpage

\begin{figure}
\epsscale{1} 
\plotone{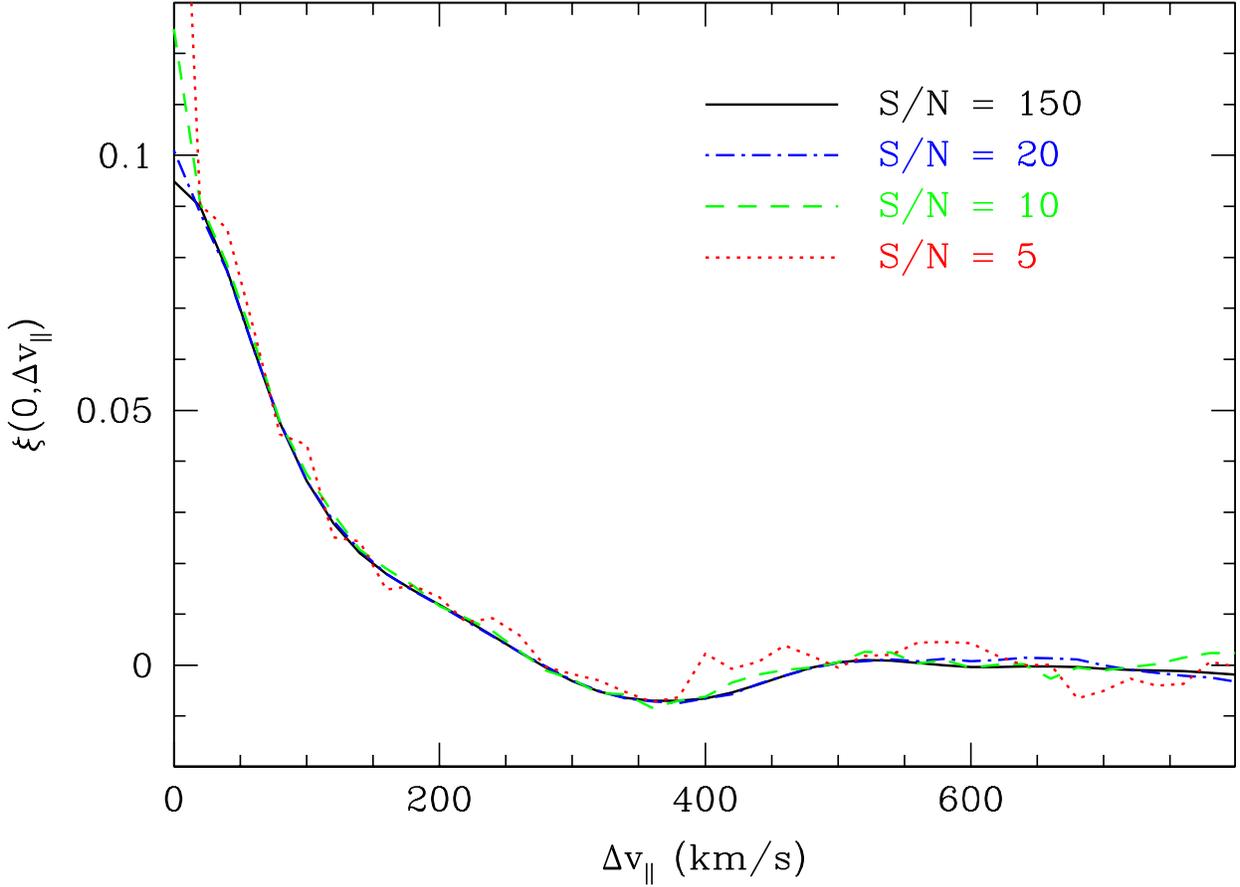} 
\figcaption{The effects of photon noise on the shape of the
autocorrelation function.  The solid line shows the autocorrelation
computed from our original ESI spectrum of Q1422 (FWHM = 55 \kms, S/N
= 150 per resolution element).  Additional lines show the
autocorrelation recomputed after adding random Gaussian noise to the
original spectrum such that the resulting S/N per resolution element
is 20 (dash-dotted line), 10 (dashed line), and 5 (dotted line).
Autocorrelation functions were computed in longitudinal velocity lag
steps of 20 \kms.  The spike at $\mdvpar = 0$ reflects only a single
point for each curve. }
\label{noise}
\end{figure}

\clearpage

\begin{figure}
\epsscale{1} 
\plotone{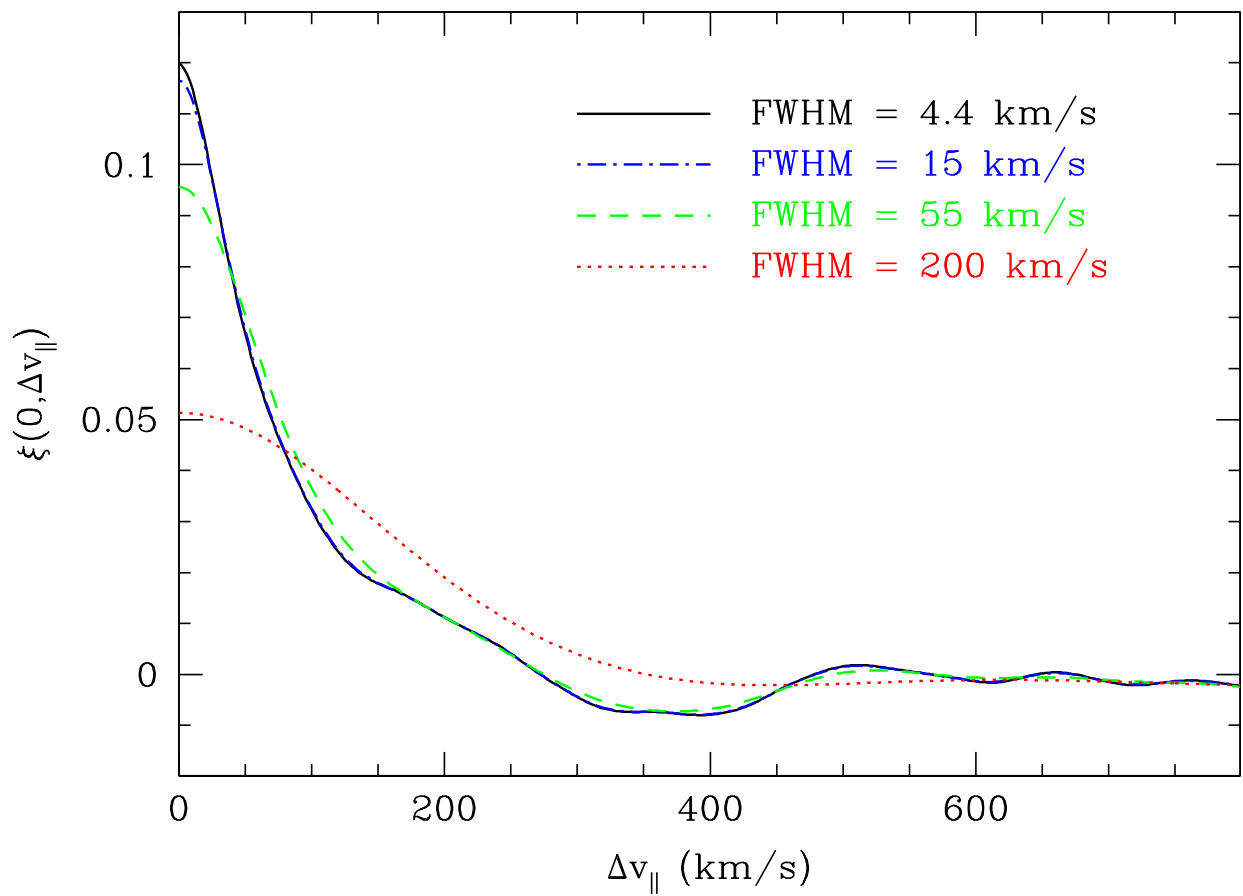}
 \figcaption{The effects of spectral resolution on the shape of the
autocorrelation function.  The solid line shows the autocorrelation
computed from our original HIRES spectrum of Q1422 (FWHM = 4.4 \kms).
Additional lines show the autocorrelation recomputed after smoothing
the HIRES data to a spectral resolution FWHM of 15 \kms~(dash-dotted
line), 55 \kms~(dashed line), and 200 \kms~(dotted line).}
\label{resolution}
\end{figure}

\clearpage

\begin{figure}
\epsscale{1} 
\plotone{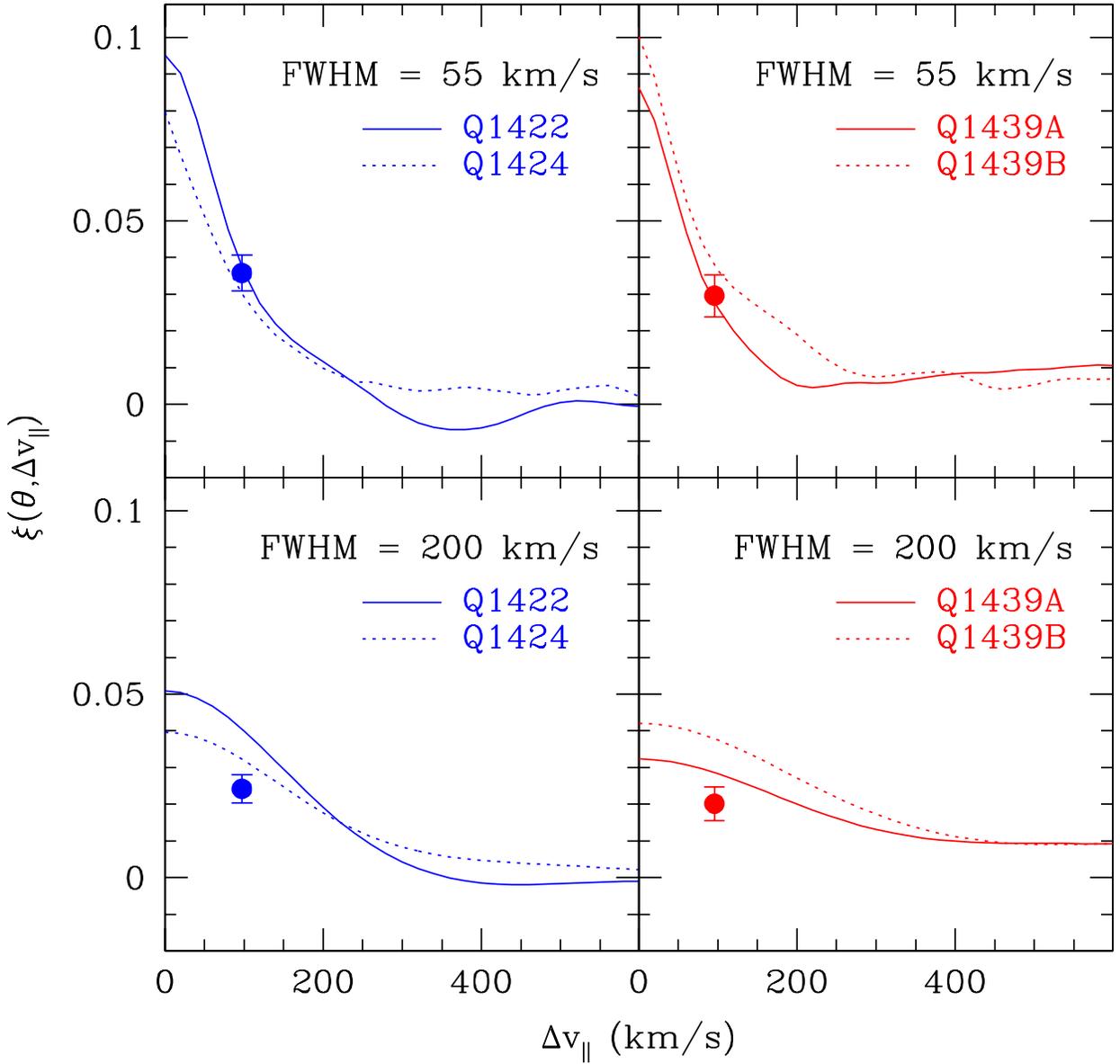} 
\figcaption{The relative effects of spectral resolution on auto- and
crosscorrelations.  {\it Top panels}: Autocorrelation functions
computed from the \lya regions in the indicated unsmoothed ESI spectra
(FWHM = 55 \kms) (solid and dotted lines) overlaid with the peak value
of the crosscorrelation computed between those spectra (filled
circles).  The peak of the crosscorrelation is plotted at a velocity
lag equal to the transverse separation between the two lines of sight
for the case of \Om = 0.3, \Ol = 0.7, and $H_0 = 70$ \kmsMpc.  {\it
Bottom panels}: Same as the top panels, here after smoothing the ESI
spectra to a resolution of 200 \kms.}
\label{auto_vs_cross}
\end{figure}

\clearpage

\begin{figure}
\epsscale{0.39}
\plotone{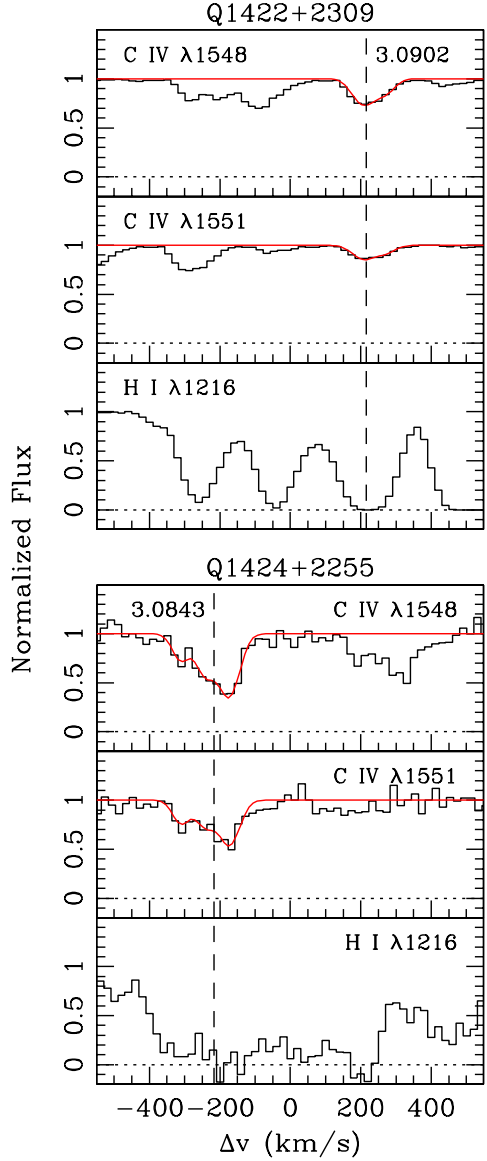}
\figcaption{{\it Top three panels:} Continuum-normalized ESI spectrum
of Q1422 (histogram) showing the \civ and \hi absorption for the
system at $z=3.0902$ reported by \citet{rauch01}.  The continuous
lines show the Voigt profile fits computed from HIRES data where the
profiles have been smoothed to ESI resolution.  {\it Bottom three
panels:} Continuum-normalized ESI spectrum of Q1424 (histogram)
showing the \civ and \hi absorption for the system measured herein at
$z=3.0843$.  The continuous lines show Voigt profile fits computed from
the ESI data.  All components in both spectra are placed on a common
velocity scale in order to demonstrate the longitudinal velocity
separation between these systems.}
\label{q1422_q1424_civ_z3.08}
\end{figure}

\clearpage

\begin{figure}
\epsscale{0.93} 
\plotone{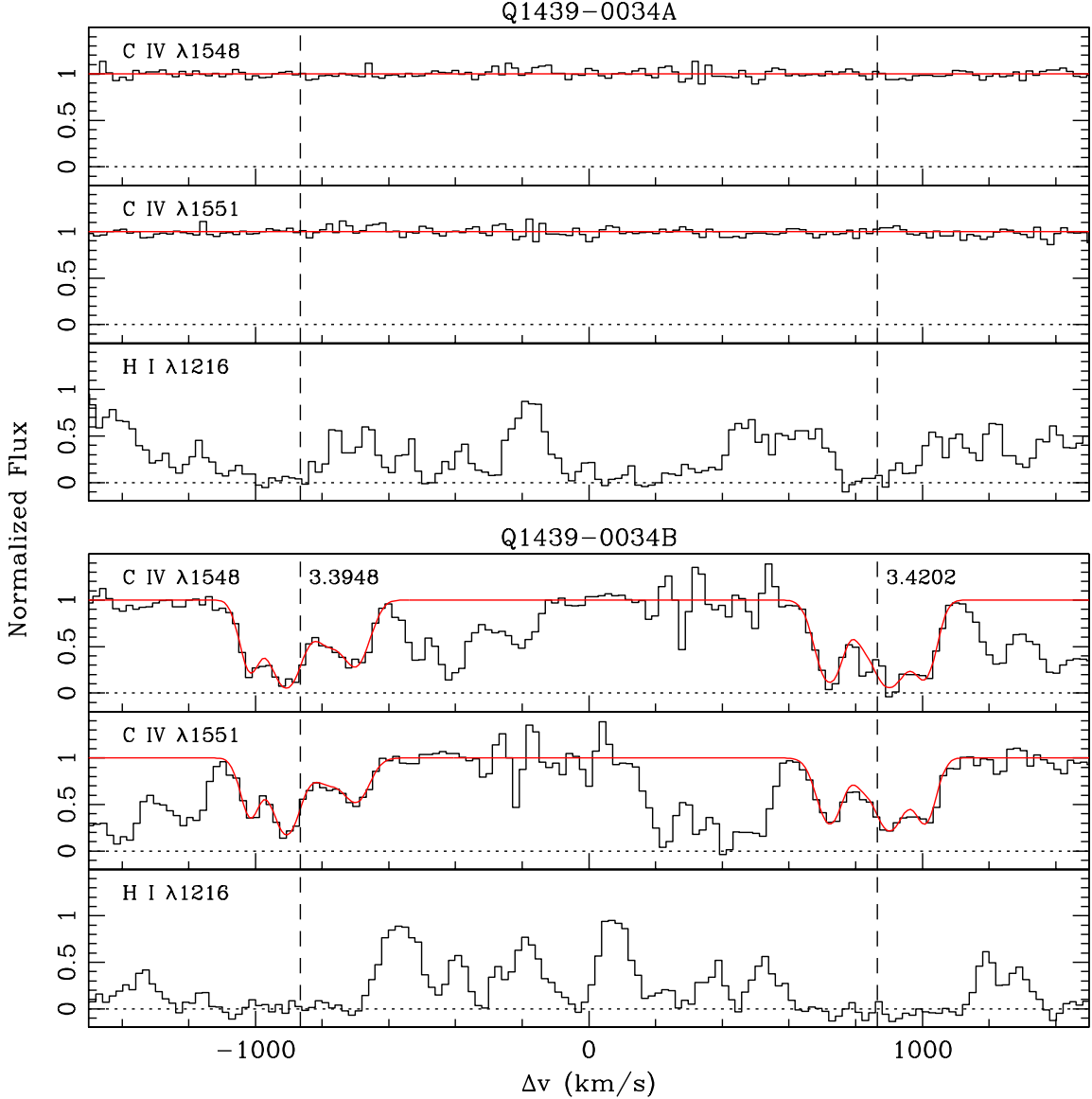}
\figcaption{{\it Top three panels:} Continuum-normalized ESI spectrum
of Q1439A (histogram) covering the \civ and \hi regions near $z =
3.4$.  The continuous lines in the top two panels demonstrate the
absence of \civ over this interval.  {\it Bottom three panels:}
Continuum-normalized spectrum of Q1439B (histogram) showing the \civ
and \hi absorption for the systems at $z = 3.3948$ and $z=3.4202$.
The continuous lines show the Voigt profile fits to the \civ
absorption.  Several of the \civ $\lambda$1548 pixels near 850 \kms~in
this plot are heavily affected by skylines and have been excluded from
the fit.  All components in both spectra are placed on a common
velocity scale in order to demonstrate the longitudinal velocity
separation between these systems.}
\label{ab_civ_z3.4}
\end{figure}

\clearpage

\begin{figure}
\epsscale{0.39} 
\plotone{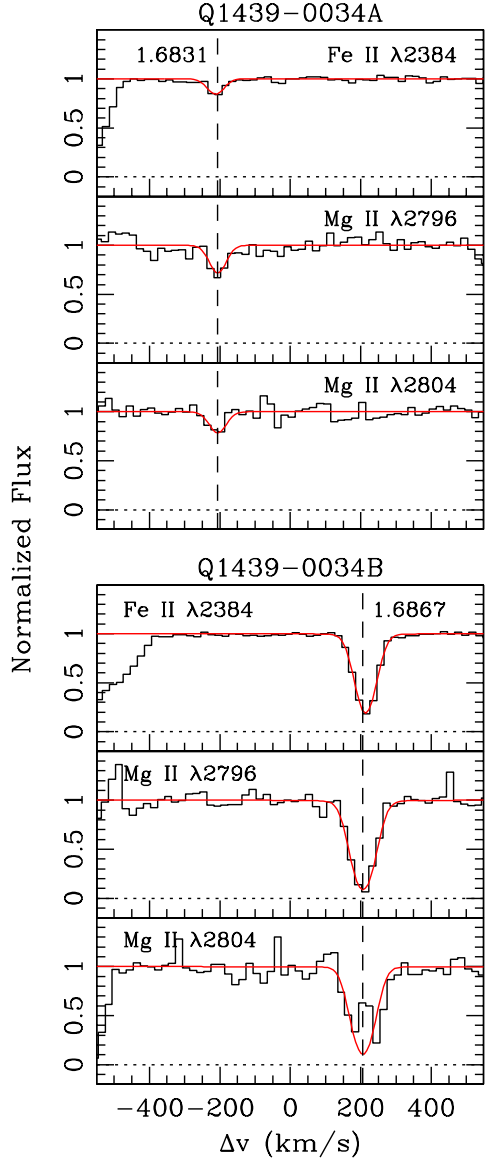} 
\figcaption{{\it Top three panels:} Continuum-normalized spectrum of
Q1439A (histogram) showing the \mgii and \feii absorption for the
system at $z=1.6831$.  The continuous lines show Voigt profile fits.
{\it Bottom three panels:} Same as the top panels, here for the system
at $z = 1.6867$ in Q1439B.  Several of the \mgii $\lambda$2804 pixels
in Q1439B near 200 \kms~in this plot are heavily affected by skylines
and have been excluded from the fit.  All components in both spectra
are placed on a common velocity scale in order to demonstrate the
longitudinal velocity separation between these systems.}
\label{ab_mgii_z1.68}
\end{figure}

\clearpage

\end{document}